\documentclass[12pt,preprint]{aastex}

\def\fxfk {F$_X$/F$_K$}
\def\nh {${\rm N_H}$}

\def\jmk{{J$-$K$_S$}}

\def\xmm{{\it XMM-Newton}}
\def\chandra{{\it Chandra}}

\def\etal {{\it et~al.\ }}

\def\sax{{\it BeppoSAX}}

\def\arcs{{\hbox{$^{\prime\prime}$}}}

\def\H0{{\rm ~km~s^{-1}~Mpc^{-1}}}

\def\lax    {${_<\atop^{\sim}}$}
\def\gax    {${_>\atop^{\sim}}$}
\def\et{{et al.}}

\shorttitle{XMM-Newton observation of Red AGN}
\shortauthors{Wilkes et al.}

\begin{document}

\title{\xmm\ observations of Red AGN}

\author{B.J.Wilkes\altaffilmark{1}, K.A.Pounds\altaffilmark{2}, G.D. Schmidt$^3$,
P.S. Smith$^3$, R.M. Cutri$^4$, H. Ghosh$^5$, 
B. Nelson$^4$, \& D.C. Hines$^6$}
\altaffiltext{1}{Harvard-Smithsonian Center for Astrophysics, Cambridge, MA 02138, USA}
\altaffiltext{2}{Department of Physics and Astronomy, University of Leicester, Leicester LE1 7RH, UK}
\altaffiltext{3}{Steward Observatory, The University of Arizona, Tucson, AZ 85721}
\altaffiltext{4}{IPAC, Caltech, MS 100-22, Pasadena, CA 91125}
\altaffiltext{5}{Ohio State University, Department of Astronomy, 4055
MacPherson Lab, 180 W. 18th St., Columbus, OH 43210-1173}
\altaffiltext{6}{Space Science Institute, 4750 Walnut Street, Suite 205,
Boulder, Colorado 80301}

\begin{abstract}

\xmm\ spectra of five red, 2MASS
AGN, selected from a sample observed by \chandra\ to be relatively
X-ray bright and to cover a range of hardness ratios,
confirm the presence of substantial absorbing material in
three sources 
with optical classifications ranging from Type 1 to
Type 2. A flat (hard), power law continuum is
observed in the other two.
The combination of X-ray absorption and broad optical emission lines
suggests either a small (nuclear) absorber or a
favored viewing angle so as to cover the X-ray source but not the broad
emission line region (BELR).
A soft excess is detected in all three Type 1 sources.  We
speculate that this may arise 
in an extended region of ionised gas, perhaps linked with the 
polarised (scattered) optical light present in these sources.
The spectral complexity revealed by \xmm\
emphasizes the limitations of the low S/N \chandra\ data. 
The new results strengthen our earlier conclusions 
that the observed X-ray continua of red AGN are unusually hard at
energies $\ga$2 keV. Their observed spectra are consistent with 
contributing significantly to the missing hard/absorbed population of the
Cosmic X-ray Background (CXRB) although their intrinsic
power law slopes are typical of broad-line (Type 1) AGN ($\Gamma
\sim 1.7$-$1.9$).
This suggests that the missing X-ray-absorbed CXRB population may
include Type 1 AGN/QSOs in addition to the Type 2 AGN generally assumed.

\end{abstract}


\section{Introduction}

The Two Micron All-Sky Survey (2MASS) has revealed a large number of
highly reddened active galaxies (AGN) not previously found in optical/UV
color-selected surveys (Cutri \etal\ 2002, Francis \etal\ 2004).
Their typically high optical polarization (Smith \etal\
2002, 2003) suggests substantial
obscuration around the nuclear energy source. \chandra\ observations show weak,
hard X-ray emission compared with normal, broad-line, low-redshift AGN,
suggesting that these, mostly broad-lined, AGN are absorbed.
Evidence for a significant population of X-ray absorbed, broad-lined AGN
is growing, both in individual sources and samples (Silverman \etal\
2005, Perola \etal\ 2004). 
Broad Absorption Line (BAL) QSOs are another well-defined subset
with this property (Green \etal\ 2001).
Notably BALQSOs also have high optical polarisation
levels. As X-ray absorbed AGN, these sources may contribute to the
X-ray-absorbed population predicted by modelling of the Cosmic X-ray
Background (CXRB, Gilli \etal\ 2001, Comastri \etal\ 1995).

The realization that obscuration plays a critical role in the classification of AGN inspired a
fundamental change in our understanding of the phenomenon. Not only does the ``Unified Scheme"
(Antonucci 1993)
provide a basis for new observations and theoretical models, but it
also implies that many AGN may
be missed in UV-excess surveys (e.g., Masci \etal\ 1999). Ideas of what comprises an AGN continue
to expand as X-ray and radio surveys find luminous, unresolved sources
that show no signs of activity at other wavelengths (e.g. Elvis \etal\
1981, Bauer \etal\ 2004), implying that the
overall number density has been significantly underestimated. Ramifications
include revisions of the fraction and types of galaxies that harbor an active nucleus, the energy
density of ionizing radiation in the young Universe, the nature of the X-ray and far-IR backgrounds,
and the importance of accretion power in the Universe as a whole.

Although IRAS provided the first significant sample of extragalactic objects in which the bulk
of the luminosity emerges as reprocessed radiation in the IR (Soifer \etal\ 1984), its
sensitivity was sufficient to catalog only the most
nearby and/or luminous AGN. The Two Micron All Sky Survey (2MASS)
has yielded a much deeper catalog of near IR-selected AGN (Cutri \etal\ 2002) by selecting sources with
$J-K_S>2$ from the high galactic latitude 2MASS Point Source Catalogue. Spectroscopic follow-up of red
candidates reveals that $\sim$75 percent are previously unidentified emission-line AGN, with $\sim$80 percent of
these showing broad optical emission lines (Type 1: Seyfert 1 and QSO), and the remainder being
narrow-line objects (Type 2: Seyfert 2, QSO 2, and LINER; Cutri \etal\ 2002). They span a redshift range
$0.1< z <2.3$ with median z$\sim$0.25. The inferred surface density is $\sim$0.5 deg$^{-2}$ brighter than 
$K_S <14.5$, higher than that of optically selected AGN at the same IR magnitudes and indicating that 2MASS
will reveal $>25,000$ such objects over the sky. The objects also have unusually high optical polarization
levels, with $\sim$10 percent showing $P>3$ percent indicating a significant contribution from scattered
light (Smith \etal\ 2002, 2003).

In a near-infrared flux-limited survey for AGN, largely unbiased by
color-selection, Francis \etal\ (2004) found that the fraction of all
galaxies harboring AGN increases with increasing \jmk\ color.  They
could not place stringent limits on the total fraction of dust-reddened
AGN at low redshift, though, because the optical colors are dominated by
host-galaxy light and because of poor statistics for the reddest
and optically-faintest sources.  Obscuration by dust may very well
account for the lower contrast of the nuclear emission at optical
wavelengths, while leaving the near-infrared nuclear colors largely
unaffected.   A better estimate of the fraction of extinguished
AGN comes from Glikman \etal\ (2004) who examined a sample of FIRST
radio sources with 2MASS near-infrared detections, but without optical
counterparts.  They concluded that this population comprises approximately
20\% of all QSOs.

ROSAT found that, while known AGN dominate the soft (0.1--2.0 keV) cosmic X-ray background (CXRB; Lehmann \etal\
2000), an additional population of heavily absorbed AGN would be required to account for the harder high-energy
spectrum (Comastri \etal\ 1995). To match both the CXRB spectrum and the observed hard
X-ray number counts of other
pre-Chandra surveys ({\it e.g.} \sax , Fiore \etal\ 2001), the X-ray absorbed
AGN population is estimated to outnumber unabsorbed
AGN by $\sim$4:1 and perhaps to increase with z (Gilli \etal\ 2001; Comastri \etal\ 2001). Although the local
ratio of Type 2 to Type 1 AGN appears to be $\sim$2--4 (Maiolino \& Rieke 1995; Huchra \& Burg 1992), a large,
thus far undiscovered population of X-ray absorbed AGN at higher z was
required by these models.

\xmm\ and \chandra\ are finding objects in sufficient numbers to explain
85$-$90\% of the CXRB at energies \lax 5 keV (Bauer \etal\ 2004,
Worsley \etal\ 2004). While these include a significant number of hard spectrum
sources, most are at low-redshift, contrary to the expectations of
the models (Gilli \etal\ 2001). Wider area surveys, which are more
able to characterize the absorbed population, are beginning to find higher
redshift absorbed sources (Fiore \etal\ 2003, Treister \etal\ 2005),
approaching the model predictions.
The hard X-ray sources correspond to both optically faint objects and
bright, nearby elliptical galaxies, as well as more traditional, broad/narrow-line
AGN (Silverman \etal\ 2004). They frequently have red continua
($1.5<$ \jmk\ $<2.5$) which, combined with their optical colors, are consistent
with the moderate amounts of gas and dust
absorption required to match the CXRB at
energies
$< 10$ keV (Compton-thin; equivalent neutral hydrogen column
density, \nh\ $\leq\ 10^{22-23}$ cm$^{-2}$). Thus, evidence is mounting that
absorbed AGN are indeed important contributors to the CXRB, although
in some cases their nuclei are only visible at X-ray wavelengths.

Given the high surface density and similarities
of the new population of 2MASS AGN to the absorbed AGN being found
in the X-ray surveys, a census of the X-ray properties of the
2MASS AGN will aid their understanding
and show whether this previously missed population
contributes significantly to the CXRB.

We have surveyed a well-defined, color-selected subset of 44
2MASS AGN (Wilkes \etal\ 2002) using the Advanced CCD Imaging
Spectrometer array (ACIS; Nousek \etal\ 1998) on \chandra . The subset
was selected to have $B-K_S>4.3$ and $K_S<13.8$, including the brightest
and reddest objects but covering sufficient parameter space to be
representative of the new population: $0< z <0.4$, 0\%$< $P$ <9.3$\% (Smith \etal\
2001) and a full range of optical class. A comparison of the
K$_S$-band to 1 keV flux density ratios (\fxfk ,
the latter computed assuming a {\it
normal} AGN X-ray spectrum; \nh=$3\times 10^{20}$ cm$^{-2}$, $\Gamma$=2)
with those of low-redshift, broad-line AGN (see Figure 1 in Wilkes \etal\
2002) demonstrates the general weakness of X-ray emission
from the 2MASS QSOs, placing them in the range measured for BAL QSOs and Sy2
galaxies. The reddest AGN in the sample
(\jmk $>2.5$) are the weakest X-ray sources, suggesting absorption by
related dust and gas affects the emission in IR and X-ray bands respectively.

The \chandra\ observations, aimed at detecting the AGN, are generally of
too low signal-to-noise to provide constraints on the X-ray
spectra and absorption beyond those of 
hardness ratios or simple power-law fits. A normal AGN X-ray
spectrum with little absorption yields a \chandra\ hardness ratio
(HR=[H-S]/[S+H], where S, H are the net counts in energy bands
0.3$-$2.5keV and 2.5$-$8keV respectively) of HR=$-0.7$ (ACIS-S, Cycle 4).
The 2MASS sample covers the range $-0.6< HR <+0.6$, consistent with
absorbing columns of \nh =$10^{21}-10^{23}$ cm$^{-2}$ for most
sources. Spectral fits for those with higher counts reinforce this conclusion
(Wilkes \etal\ 2004). Two 2MASS AGN are consistent with no absorption
above the Galactic column ($HR <-0.5$). 
Contrary to a simple absorption
scenario in which a single screen of gas and dust obscures all continuum
emitting regions, there is no correlation between HR and \jmk, {\it i.e.} while
being X-ray-weak, the reddest objects are not necessarily those with the
hardest X-ray emission.
However, AGN
X-ray spectra often contain a number of components which modify the power
law continuum emission believed to originate in the central AGN. In 
addition to cold absorbing material, these may include: partial covering of the
source, an ionized absorber, soft excess emission, and scattering/reflection
by neutral and/or ionized material.
Such complexity can strongly affect the interpretation of a simple
hardness ratio determined from the low S/N \chandra\ spectra.

Correction of the X-ray flux for absorption based on the hardness ratios
moves only about half of the 2MASS sample into the X-ray faint region occupied by the
Elvis \etal\ (1994) broad-line AGN in \fxfk\
(Wilkes \etal\ 2004), or even the X-ray weaker, optically selected
sample of Laor \etal\ (1997). Possible explanations for this X-ray weakness
include: a high K$_S$ flux due
either to increased hot dust emission or the inclusion of extended, host galaxy
emission in these mostly low-redshift sources; 
instrincally faint and hard X-ray emission; or complex X-ray spectra that
result in our underestimating their intrinsic X-ray flux.

With the aim of removing these ambiguities for at least a subset of the \chandra\ sample
of 2MASS AGN, we recently
obtained \xmm\ spectra of 5 representative sources (Table~\ref{tb:obs}). 
Improved X-ray spectra will clarify
the relationship with other AGN properties such as optical
and IR color, optical polarisation and X-ray to IR ratio. Understanding the
individual sources is a key to determining the relation of these new, red, X-ray absorbed,
broad-lined AGN to the remainder of the AGN population.

Throughout this paper, we assume H$_{\circ}$=70 km s$^{-1}$
Mpc$^{-1}$, $\Omega_{\Lambda}$=0.7, and $\Omega_{\rm{M}}$=0.3.

\section{Observations}

Details of the \xmm\ observations of the 2MASS sources are given in Table~\ref{tb:obs}.
X-ray data were available throughout each observation from the EPIC
pn (Str\"{u}der \et\ 2001) and MOS (Turner \et\ 2001) cameras, providing
moderate resolution spectra (35 eV at Al-K (1.5 keV) and 61 eV at Mn-K (5.9 Kev)
over the energy band $\sim$0.3--10 keV. 

The X-ray data were first screened with the XMM SAS v5.4 software and events corresponding to patterns
0-4 (single and double pixel events) were selected for the pn data and patterns 0-12 for MOS1 and MOS2,
the latter then being combined. A low energy cut of 300 eV was applied to all X-ray data and known hot or
bad pixels were removed. Source counts were obtained from a circular region of 45\arcs\ radius centered on
the target source, with the background being taken from a similar region offset from, but close to, the
source. Unfortunately the background was high during parts of each observation, and these data have been
excluded from the subsequent spectral analysis, using the recommended cut-off background rates of 1
s$^{-1}$ (pn camera) and 0.35 s$^{-1}$ (MOS camera). The resulting exposures and source counts are listed
in Table 1. The integrated  data set for each source was then available for spectral analysis. Individual
EPIC spectra were binned to a minimum of 20 counts per bin to facilitate use of the $\chi^2$
minimisation technique in spectral fitting. Spectral fitting was based on the Xspec package (Arnaud
1996) and all fits included absorption due to the relevant line-of-sight Galactic column. Errors are
quoted at the 90\% confidence level ($\Delta \chi^{2}=2.7$ for one interesting parameter).

\section{X-ray spectral analysis} 
The optical and \chandra\ X-ray properties of the 5 AGN are listed in Table~\ref{tb:Xopt}
and repeated here as we describe the X-ray spectral fitting of the
\xmm\ data. Note that the \chandra\ fits were made over the 0.3-8 keV band, with
a power law plus cold absorption (using the CIAO software, Fruscione \&
Siemiginowska 2000). In assessing the \xmm\ spectra, we first fitted
a power law to the energies $\ga$2 keV. This is the conventional approach
in analysing AGN X-ray spectra, generally to minimise the effects of absorption. More particularly, in
this study, that approach allows a direct comparison of individual source spectra with the CXRB which has the form of a 
power law of photon index $\Gamma$=1.4$\pm0.1$ above $\sim$2 keV. The pn and MOS
data are fitted simultaneously in each case, with only the power law index and normalisation untied.

\subsection{2MASS J09184860+2117170}
In the optical this source shows an intermediate (1.5) spectral classification
and lies at a redshift of
z=0.149. The Galactic column is \nh\ = 4.1$\times 10^{20}$ cm$^{-2}$. It is a highly reddened 
object in the \chandra\ sample, with
\jmk =2.23, but the \chandra\ data are soft (HR=$-$0.6) and
a simple broad-band power law fit yields a power law index
close to that for a normal, unabsorbed Seyfert 1 galaxy (Table~\ref{tb:Xopt}).
Thus, if absorption is responsible for the red color, X-ray spectral
complexity must be present causing the
X-ray absorption to be underestimated in a simple power law fit.

\xmm\ observed this source to be a factor $\sim 4-$5 fainter than
during the \chandra\ observation. 
Fitting the pn and MOS data to a power law above 2 keV yielded flat
but poorly constrained spectral indices of $\Gamma$=1.26$\pm$0.39 (pn) and
$\Gamma$=1.21$\pm$0.42 (MOS). Extrapolating this power law to 0.3 keV was clearly a poor fit ($\chi^2$=59 for 
25 degrees of freedom) with a soft excess and also some upward curvature in the highest energy channels (Figure 1a). 
Since the latter
was suggestive of reflection a further power law fit was made, now including the data down to 1 keV (to tighten the
constraints) and adding a cold reflection component modelled in Xspec by PEXRAV (Magdziarz \&
Zdziarski 1995).
The best fit had a high reflection component, with R$\sim$6. The power law index increased by $\sim$0.3 in this fit,
but the soft excess remained. Adding a blackbody then provided an acceptable fit over the whole 0.3-8 keV energy band.
Interestingly, a gaussian emission `line' matched the soft excess rather better and this is included in the best-fit 
model ($\chi^2$=13 for 
20 degrees of freedom) reproduced in Figure 1b. 
The parameters of the best-fit model are an underlying power law index $\Gamma$=1.65$\pm$0.33 (pn) and
$\Gamma$=1.56$\pm$0.38 (MOS), with cold reflection (R$\sim 5$,
EW(FeK$\alpha$)$\sim 0.5$ keV), and a gaussian emission line at 0.44$\pm$0.13 keV, 
with line width $\sigma \sim$75 eV and
flux $\sim$$4\times 10^{-5}$ photon cm$^{-2}$s$^{-1}$ (Table~\ref{tb:xmm}).
This `line' is a rough approximation to the blend of line emission
expected from ionised gas. More detailed modeling of this feature,
unwarranted in the current low S/N data, would require an accounting for
the uncertain energy calibration below E$\sim 0.5$ keV.\footnote{As
discussed in \xmm\ document XMM-SOC-CAL-TN-0018 by M. Kirsch (2004).}

We determine an upper limit on the X-ray column density,
\nh $\la 1.1\times 10^{21}$ cm$^{-2}$
(90\% confidence). Including absorption at this level in the spectral
fit steepens the deduced power law to $\Gamma \sim 1.8$, well within the 
normal range.

The observed energy fluxes (pn camera) (Table~\ref{tb:flux})
were $4.1\times 10^{-14}$ erg cm$^{-2}$s$^{-1}$ (0.3-1 keV), $3.2\times 10^{-14}$ 
erg cm$^{-2}$s$^{-1}$ (1-2 keV), and $1.9\times 10^{-13}$ erg cm$^{-2}$s$^{-1}$ (2-10 keV).
The overall 0.3-10 keV luminosity was $1.4\times 10^{43}$ erg s$^{-1}$ with $6\times 10^{41}$ erg s$^{-1}$ in the gaussian
soft X-ray emission component.

\subsection{2MASS J10514425+3539306}
At a redshift of z=0.158 this source has been optically classified as a
type 1.9 QSO (Smith \etal\ 2003). The Galactic column is 
\nh\ = 1.9$\times 10^{20}$ cm$^{-2}$. Although less strongly reddened than
2MASS J09184860+2117170,
it had a
harder HR=$-$0.2 in the \chandra\ observation. The \chandra\ data
had sufficient counts to determine a power law fit of
$\Gamma$=1.6$\pm$0.2 attenuated by a column of \nh\ = (5.6$\pm$1.4)$\times 10^{21}$ cm$^{-2}$. 

The \xmm\ data confirmed and further constrained the \chandra\ spectral fit. A power law
for energies $\ga$2 keV yielded photon indices
$\Gamma$=1.67$\pm$0.11 (pn) and $\Gamma$=1.51$\pm$0.11 (MOS).
When extrapolated to a lower energy strong attenuation is evident (Figure 2a). Refitting the data
over the 0.3--9 keV band with a power law plus absorber (modelled by the
photoionised absorber model, ABSORI in Xspec) produced an
excellent fit ($\chi^2$=103 for 115 degrees of freedom), with an absorbing column (solar abundances) of cold matter
($\xi$$\leq$0.7) of
\nh\ = (7.2$\pm$0.7)$\times 10^{21}$ cm$^{-2}$. The power law index steepened in this broad-band fit (reproduced
in Figure 2b) to $\Gamma$=1.79$\pm$0.08 (pn) and $\Gamma$=1.70$\pm$0.08 (MOS). 

The observed fluxes (pn camera) (Table~\ref{tb:flux})
were $5\times 10^{-14}$ erg cm$^{-2}$s$^{-1}$ (0.3-1 keV), $4\times 10^{-13}$ 
erg cm$^{-2}$s$^{-1}$ (1-2 keV) and $2.0\times 10^{-12}$ erg cm$^{-2}$s$^{-1}$ (2-10 keV).
The observed 0.3-10 keV 
luminosity was $1.5\times 10^{44}$ erg s$^{-1}$, and $2.3\times 10^{44}$ erg s$^{-1}$ with the absorbing column removed.
Such a high luminosity supports the identification of
2MASS J10514425+3539306 as a QSO, rather than a LINER/starburst galaxy. In
addition, detailed comparison of the X-ray fluxes
shows the source had brightened by a factor of $\sim$1.5--2 between the \chandra\ and \xmm\ observations, larger than 
the current uncertainties in
cross-calibaration.

\subsection{2MASS J13000534+1632149}

This object is optically classified as a Type 2 QSO (Schmidt
\etal\ 2002, Smith \etal\ 2003), updated
from the 1.x classification in Smith \etal\ (2002), which was based on a
low S/N discovery spectrum. 
It lies at a redshift of z=0.08 and the Galactic column is \nh\ = 2.0$\times 10^{20}$
 cm$^{-2}$. \chandra\  found this source to have the hardest
spectrum of our sub-set, with HR=0.2 (Wilkes \et\
2002), $\Gamma =1.5$ and \nh $\sim 1.8 \times 10^{22}$ cm$^{-2}$ (Table~\ref{tb:Xopt}).
The \xmm\ observation found an X-ray flux level essentially unchanged from the
\chandra\ value ($\leq$20\%\ change based on the predicted vs. observed
\xmm\ count rates).

A power law fit for energies \gax 2 keV yielded very flat (hard) photon
indices of $\Gamma$=1.16$\pm$0.16 (pn)
and $\Gamma$=1.1$\pm$0.2 (MOS), with an extrapolation to lower energies showing a cut-off similar to that in
2MASS J10514425+3539306 (Figure 3a). Adding ABSORI to model the attenuation, the best-fit power law
steepened to $\Gamma$=1.67$\pm$0.20 (pn) and $\Gamma$=1.65$\pm$0.21 (MOS), with a large absorbing column 
(solar abundances) of \nh\ = (2.9$\pm$0.5)$\times 10^{22}$ cm$^{-2}$. The ionisation
parameter ($\xi$= $L/nr^2$)
was $\xi$=4.1$\pm$1.7,
indicating the absorbing material is, at most, only weakly ionised.

Although the absorbed power law fit was statistically very good ($\chi^2$=20/28), the residuals 
showed an excess near 6 keV. Adding a narrow gaussian line to the model further improved the fit, to 
$\chi^{2}$=11/24, with a line energy (AGN rest-frame) of 6.20$\pm$0.35 keV and EW$\sim$250 eV.
This addition to the model resulted in a further steepening of the power law index by $\sim$0.1.
Figure 3b illustrates the best-fit pn spectrum.

The observed fluxes (pn camera) (Table~\ref{tb:flux})
were $1.4\times 10^{-14}$ erg cm$^{-2}$s$^{-1}$ (0.3-1 keV), $1.45\times 10^{-13}$ 
erg cm$^{-2}$s$^{-1}$ (1-2 keV) and $2.35\times 10^{-12}$ erg cm$^{-2}$s$^{-1}$ (2-10 keV).
The observed 0.3-10 keV 
luminosity was $3.7\times 10^{43}$ erg s$^{-1}$, and $6.8\times 10^{43}$ erg s$^{-1}$ with the absorbing column removed.

\subsection{2MASS J14025120+2631175}
The final two objects in our sample are optically classified as Type 1 QSOs.
2MASS J14025120+2631175 is
at a redshift of z=0.187, and the Galactic column \nh\ = 1.4$\times 10^{20}$ cm$^{-2}$. The \chandra\ observation 
yielded a hardness ratio HR=$-$0.6 indicating significant
X-ray flux in the 0.5-2.5 keV band. The simple power law fit (Table~\ref{tb:Xopt})
showed a steep power law of 
$\Gamma$=2$\pm$0.4, with no significant absorption.

Our \xmm\ observation found 2MASS J14025120+2631175 to be the brightest X-ray source in the sample, at a flux
level consistent with that observed by \chandra.
A power-law fit for energies \gax 2 keV yielded moderately hard spectral
indices $\Gamma$=1.53$\pm$0.16 (pn) and $\Gamma$=1.4$\pm$0.15 (MOS).
Extrapolating this fit to 0.3 keV revealed a soft excess (Figure 4a). Modelling the soft
excess with a gaussian emission line (again a marginally better fit than a blackbody; $\Delta \chi^{2}$= $-4$ for $-1$ dof),
gave an excellent fit ($\chi^2$=110/102, Figure 4b). The `line' energy
(AGN rest frame) was 0.44$\pm$0.10 keV, with a width $\sigma$$\sim$100 eV and
flux $\sim$$4\times 10^{-4}$ ph
cm$^{-2}$s$^{-1}$. 
We also determine an upper limit on the X-ray column density,
\nh $\la 4\times 10^{20}$ cm$^{-2}$
(90\% confidence); including absorption at this level in the spectral
fit steepens the deduced power law to $\Gamma \sim 1.6$, closer to the
normal range for type 1 AGN.

The observed fluxes (pn camera) (Table~\ref{tb:flux})
were $7.5\times 10^{-13}$ erg cm$^{-2}$s$^{-1}$ (0.3-1 keV), $4.7\times 10^{-13}$ 
erg cm$^{-2}$s$^{-1}$ (1-2 keV) and $1.9\times 10^{-12}$ erg cm$^{-2}$s$^{-1}$ (2-10 keV).
The overall 0.3-10 keV luminosity was $2.6\times 10^{44}$ erg s$^{-1}$ with $1.6\times 10^{43}$ erg s$^{-1}$ in the gaussian
soft X-ray emission component.

\subsection{2MASS J23444958+1221432}
2MASS J23444958+1221432 is classified as a Type 1 QSO, lies at a redshift of z=0.199, and is viewed through a Galactic column
of \nh\ = 4.7$\times 10^{20}$ cm$^{-2}$. It had a hardness ratio,
HR=$-0.4$ in the \chandra\ observation, with a simple power law fit finding a power law of 
$\Gamma$=2$\pm$0.4, plus an absorbing column of \nh\ = (4$\pm$2)$\times 10^{21}$
cm$^{-2}$.

Our \xmm\ observation found 2MASS J23444958+1221432 to be substantially fainter than during the \chandra\ observation, 
by a factor of $\sim$2--3.
Proceeding as before we first fitted a power law to the data above $\sim$2 keV, obtaining 
typical Type 1 indices of $\Gamma$=1.70$\pm$0.16 (pn) and $\Gamma$=1.58$\pm$0.18 (MOS). Extrapolating this
fit to 0.3 keV showed the spectrum to be complex (Figure 5a) with evidence of absorption and
also possibly a soft excess. 

To first model the absorption, we again added ABSORI to the power law, finding a reasonably good fit
($\chi^2$=91/73) with a column density \nh\ =(9.7$\pm$1.5)$\times 10^{21}$ cm$^{-2}$ of moderately ionised gas. 
The ionisation parameter $\xi$=21$\pm$9
is constrained by the energy of the flux upturn, observed below $\sim$0.7 keV, which 
corresponds primarily to the absorption edge of ionised OVII in the model. However, visual examination of this
spectral fit (Figure 5b) shows the absorbed power law model fails to match fully
the observed spectral upturn.

We therefore added an emission component,
modelled by a gaussian line, obtaining a significantly improved fit
of $\chi^2$=65/69. As with 2M0918, this `line' approximates the blend
of emission lines from ionised gas and accurate modeling would require
accounting for the uncertain energy calibration at E$<$0.5 keV.
In this fit the continuum curvature below $\sim$2 keV was again modelled by ABSORI, while
the low energy upturn was due mainly to the gaussian line. A consequence was that the ionisation parameter of
the absorbing column fell to $\xi$=0.02$\pm$0.2. Figure 6 reproduces this `best' spectral model for
2MASS J23444958+1221432. The power-law indices in this fit were $\Gamma$=1.84$\pm$0.27 (pn) and $\Gamma$=1.91$\pm$0.22
(MOS). The low ionisation absorber had \nh\ =(6.5$\pm$2.6)$\times 10^{21}$cm$^{-2}$. The  gaussian line had an
energy (AGN rest frame) of 0.64$\pm$0.05 keV and flux of $10^{-4}$ ph cm$^{-2}$ s$^{-1}$
(Table~\ref{tb:xmm}).

The observed fluxes derived from this best fit (pn camera) (Table~\ref{tb:flux})
were $9.0\times 10^{-14}$ erg cm$^{-2}$s$^{-1}$ (0.3-1 keV), 
$1.4\times 10^{-13}$ 
erg cm$^{-2}$s$^{-1}$ (1-2 keV) and $5.1\times 10^{-13}$ erg cm$^{-2}$s$^{-1}$ (2-10 keV).
The overall 0.3-10 keV luminosity was $7.9\times 10^{43}$ erg s$^{-1}$ with $2.9\times 10^{42}$ erg s$^{-1}$ in the gaussian
soft X-ray emission component.

\section{Discussion}


The primary motivation for the \xmm\ observations was to study the X-ray
spectral complexity to estimate its effect on the X-ray properties of
these sources in the lower S/N \chandra\ data and thus guide our
assessment of the larger \chandra\ sample.
Although the \xmm\ exposures were typically a factor $\sim$2 shorter than
planned, due to the rejection of noisy data, the recorded counts were greater
and hence the broad-band X-ray spectra were better determined than in
the \chandra\ survey.
For simplicity we refer to the
individual AGN using only their first 4 numbers,
{\it e.g.} 2MASS J091848+2117 is 2M0918, in the following discussion.
Since the spectral form is known to change with flux
level in AGN, it is important to note that 2M1051 was
brighter, while 2M0918 and 2M2344 were fainter in the \xmm\ observations. 

\subsection{The X-ray Properties of the 2MASS AGN.}

Our \xmm\ observations showed two of our five sources 
to have extremely hard observed spectra ($\Gamma$ \lax 1.3, Table~\ref{tb:xmm})
above 2 keV (2.16-2.37 keV adjusted for redshift). In the case
of 2M0918 our modelling suggests this is due to a strong (non-varying)
reflection component, consistent with the low flux state during
the \xmm\ observation, while 2M1300 exhibits the largest absorbing column, and
the only absorber with significant opacity above 2 keV. Allowing for a large
reflection component in 2M0918 and the observed cold absorption in 2M1300 shows
the underlying power law slopes to be `normal' for a Type 1
QSO (Table~\ref{tb:xmm}). These two cases show how the
presence of strong reflection or a Compton thin absorber can make the observed
spectrum hard above $\sim$2 keV.
In two other cases (2M1051 and 2M2344) the \xmm\ spectra show substantial
cold absorption, though not sufficient to
greatly affect the spectral slope above 2 keV. 
For none of these four red AGN does the intrinsic power law continuum appear
unusually hard.
The fifth, 2M1402, has a somewhat harder than average power law ($\Gamma =1.54$)
and a soft excess but no detected absorption in the \xmm\ data.

A soft X-ray excess is found in all three Type 1 sources
({\it i.e.} Type 1$-$1.5: 2M0918, 2M1402 and 2M2344),
in each case having a luminosity $\sim$5\%\ of the corresponding 2--10 keV
luminosity. All three soft excesses are well modelled by a gaussian emission
line centred at 0.55$\pm 0.1$ keV (in the AGN rest frame), providing a rough
approximation to the blend of emission lines expected from ionised gas.

\subsubsection{Comparison with the \chandra\ Data}

The \chandra\ X-ray parameters for the five AGN are listed,
along with other multi-wavelength properties, in Table~\ref{tb:Xopt}.
The \chandra\ values are generally poorly constrained.
\xmm\ detects complexity in three sources, two of which appeared `normal' in
the lower S/N \chandra\ data, and variability in three.
\xmm\ also informs us that a simple hardness ratio is an extremely
limited indicator of an AGN spectrum. In particular, using a
hardness ratio to deduce the presence and amount of X-ray absorption
can be highly misleading or completely wrong. 
Spectral hardness may be due
either to absorption, as originally assumed (Wilkes \et\ 2002),
or a hard power law or
strong reflection component, similar to those of Seyfert 2
galaxies (Turner \etal\ 1997b),
or hardness may be masked by a soft excess component.

The spectral parameters of the two Type 2 ({\it i.e.} Type 1.8$-$2)
sources, 2M1051 and 2M1300, 
are consistent between the two observations, despite a  $\sim
\times 2$ lower flux level for the latter.

The \xmm\ spectrum of 2M0918 (Type 1.5) is different from that seen by
\chandra\ (Table~\ref{tb:xmm}, Table~\ref{tb:Xopt}).
To make a direct comparison, we determined a flux-based
hardness ratio (HR$_F$) from the observed flux in the
soft (0.3$-$2.5 keV) and hard (2.5$-$8.0 keV) bands using the 
best fit spectrum in each case. A ``normal" power law ($\Gamma = 1.9$)
with a typical Galactic \nh (3$\times 10^{20}$ cm$^{-2}$) yields a
HR$_F$=$-$0.14 using this method. For 2M0918, the
\chandra\ data give HR$_F$=$-$0.12 (assuming no
detected absorption) while for \xmm\ the value is significantly harder,
HR$_F$=0.21. The error on these values is $\sim 0.1$ due to the
uncertainty in the spectral fit. 
The explanation may lie in the X-ray flux being a factor $\sim$4--5
fainter in the \xmm\
observation, deduced by comparing the predicted and actual \xmm\ count rates.
A variable power-law component may have decreased in flux
allowing the harder (reflection-dominated?) power law continuum to
dominate, while also
causing the soft excess to be more visible if the latter is
intrinsically less variable. The change in spectrum was confirmed
by fitting the \chandra\ data with the best fit \xmm\ spectral
parameters, which gave a significantly poorer fit than that listed in
Table~\ref{tb:Xopt} ($\Delta \chi^2$ = 3.4). 

The flux levels and spectral parameters for Type 1 2M1402,
are consistent, although \xmm\ reveals a
soft excess that was not visible in the \chandra\ data.
The flux-based
hardness ratio is harder in the \xmm\ data (HR$_F = -$0.04 compared with
HR$_F = -$0.23)
suggesting some variation.

Type 1 2M2344, similarly to 2M1402, has consistent spectral parameters
but \xmm\ reveals a soft excess which also masks a somewhat higher
absorbing column density than is visible in the \chandra\ data.
While the \chandra\ data, with only $\sim 250$ counts, do not detect the
soft excess, fitting these data with the \xmm -deduced parameters yields
a $\chi ^2$ similar to that for the power law fit.
Thus, despite a factor $\sim \times 2$ reduction in flux in the \xmm\ data,
there is no evidence for spectral variability between the two observations.

Given the variety of behavior in these five AGN, comparison of
the \chandra -deduced \nh\ and flux values with those from the \xmm\ data
(Table~\ref{tb:xmm}) shows no {\bf systematic} effects 
which could be applied to the full \chandra\ sample. While this
validates any statistical results derived from the \chandra\ data, it
also emphasizes that the deduced properties of individual AGN are not reliable
and that the resulting high error levels will generate significant scatter.

\subsection{The relation between X-ray spectral and multi-wavelength properties.}
Although this sample comprises only five AGN, it is instructive to
investigate any systematic relation between the various X-ray
properties and their multi-wavelength properties.

Recent X-ray surveys have demonstrated
that the X-ray hardness seen at fainter flux levels
is generally due to absorption
(Kim \etal\ 2004). The relation between optical$-$infra-red colors
and X-ray hardness (Mainieri \etal\ 2002) suggests that the obscuring
material includes dust which then reddens the optical/near-IR continuum,
although the dust column density is generally 
lower than expected based on a Galactic gas-to-dust ratio
(Alonso-Herrero, Ward and Kotilainen 1997, Maiolino \etal\ 1998).
Although there are exceptions, both in individual AGN and subsets of the
popluation such as
Broad Absorption Line (BAL) QSOs and the Red AGN reported here,
X-ray or optically selected AGN samples show
that, statistically, broad-lined AGN are soft (presumed unabsorbed) in the
X-ray while narrow-lined AGN are hard (presumed absorbed)
(Turner \etal 1998, Risaliti
\etal\ 1999, Laor \etal\ 1997).

The two Type 2 QSOs in our sample, 2M1051 and 2M1300, with optical
types 1.9 and 2.0 respectively, are straight-forward.
The X-ray spectral shape is typical of AGN, $\Gamma \sim 1.8$,
and their X-ray absorption (\nh $\sim 10^{22}$ cm$^{-2}$,
Table~\ref{tb:xmm}) is
at the low end of the observed range for optical type 2 AGN
(Risaliti \etal\ 1999). Both have significant polarized
optical flux, particularly when dilution by host galaxy light is included
(Table~\ref{tb:Xopt}, Schmidt \etal\ 2003)
and show broad lines in their
polarized optical flux spectra. The FWHM of H$\alpha$
in 2M1300 is unusually broad, $\sim 18 000$ kms$^{-1}$ (Schmidt \etal\
2002). Interpretation as edge-on sources viewed through a
significant column density of obscuring material, and for which the direct
nuclear light is visible due to a mirror above/below the plane of the
obscuring material, is standard in the unified scheme (Antonucci 1993).
Any X-ray complexity is hidden in this edge-on view.

However, the three Type 1 AGN in this sample all show different behavior
which does not imediately fit into a simple unification scenario.
While the B$-$R colors follow the expected trend of increased reddening
as the type progresses from 1 to 2 (Table~\ref{tb:Xopt}, Wilkes \et\ 2004),
the \jmk\ color appears unrelated.
For example, 2M0918, which has no detected X-ray absorption, is
the reddest (\jmk\ $\ga$ 2.2). Possibly the 
underlying relation between X-ray absorption and/or
optical type with \jmk\ colors is masked by the wide intrinsic range
of \jmk, $\pm 0.5$ (full range) in Type 1.

2M1402 is a Type 1 QSO with no detected X-ray absorption, normal for a
Type 1 source except for its unusually red \jmk\ color. If we assume no
absorption is present, the red color could be explained by a large
amount of hot dust increasing the K$_S$ emission.

2M2344, also a Type 1, has significant X-ray absorption. The obscuration
must cover the X-ray source but not the BELR, requiring either 
a small, dense absorber interior to the BELR or a specific orientation
with a line-of-sight passing through the wind above a disk/torus
(Konigl \& Kartje 1994) which allows a view of the BELR.

2M0918 is the most difficult to understand. It is a Type 1.5 in which the
broad emission lines are partially obscured. The X-ray continuum is
hard above 2 keV but no absorption is detected (\nh $\la 1.1 \times
10^{21}$ cm$^{-2}$).
The optical and IR continua are redder than is
typical of Type 1 AGN, the former is consistent with A$_V \sim 2$ (\nh
$\sim 3.4\times 10^{21}$ cm$^{-2}$ assuming a normal gas-to-dust ratio and
that the Elvis \etal\ (1994) median SED is unabsorbed).
This is higher than the maximum column density of cold absorbing material
indicated by the \xmm\ data, which is thus insufficient to explain the
optical$-$IR continuum unless the gas-to-dust ratio is unusually low. 
A higher S/N \xmm\
observation has been requested to check the level of X-ray absorption.

Alternatively, a significant contribution by scattered light,
as in the classic case of NGC1068 (Turner \etal\ 1997a), 
could result in the X-ray emission being dominated by the
unabsorbed light from the nuclear regions, scattered off the mirror
responsible for the high polarization of the optical emission.
This would imply strong absorption, to suppress the direct continuum
emission, and neither emission nor absorption would be
detectable in the \chandra\
or \xmm\ energy bands.
In this case, the $[$OIII]$\lambda$5007 line flux, which correlates
strongly with the intrinsic X-ray flux in AGN (Mulchaey \etal\ 1994,
Turner \etal\ 1997a),
would appear abnormally strong relative to the observed X-ray flux
and the X-ray luminosity would be unusually low.
However, the F($[$OIII])/F$_X$ ratios for the \xmm -observed
AGN are within $\sim 2 \sigma$ of the
the mean value of log F$_{[OIII]}\/$/F$_X$ = $-1.89\pm0.25$ for Sy1 galaxies
(Mulchaey \etal\ 1994), for 2M0918 the value is $-$1.49 (Wilkes \etal\ 2004).
In comparison, the value for NGC1068, for which the X-ray emission is
known to be purely due to scattered light, the value is $\sim$0.6. 
In addition, the X-ray luminosities of the AGN in our sample are $\sim 10^{44}$
erg s$^{-1}$, in the normal range for AGN (Table~\ref{tb:xmm}).
It is thus unlikely that
we are primarily observing scattered light in the X-rays.

In summary, all three Type 1 can be explained by a unification model
which includes lower column density material above/below the disk/torus,
such as those in disk-wind models (Konigl \& Kartje 1994, Murray \&
Chiang 1995, Elvis 2000).
However additional emission and absorption from dust must also be present.
Alternatives, such as a small, dense absorber, can describe individual
sources, but not all three using the same scenario.

\subsection{2MASS AGN and the CXRB}

Models of the CXRB require a significant population of X-ray absorbed AGN
in order to match the CXRB spectrum which is significantly flatter than
that of unabsorbed AGN (Comastri \etal\ 1995, Gilli \etal\ 2001).
This population has generally been assumed to
be dominated by Type 2 AGN and QSOs whose X-ray emission is generally absorbed
(Turner \etal\ 1998).
The observations have not yet found this expected population at high
redshift (Bauer \etal\ 2004).
However, the combination of X-ray hardness and visible optical
broad-lines in red AGN suggests that a subset of the broad-line AGN
population may also be significant contributors.
The \xmm\ results confirm and expand on this
possibility, demonstrating that the observed X-ray spectrum may be hard at
energies above 2 keV, even where \chandra\ did NOT detect this hardness (2M0918,
2M1402). All five sources are affected by absorption and/or a soft emission
component at energies below $\sim 2$ keV.
For 2M0918 the \xmm\ data cannot distinguish between an
intrinsically hard power law and a strong reflection component which
flattens the X-ray spectrum at higher energies.

Interestingly, accounting for spectral complexity
in our diverse sample may explain why AGN with apparently normal,
steep continua ($\Gamma \sim 1.7-1.9$, Nandra \& Pounds 1994)
can more closely
match the CXRB in the (observed) 2--10 keV band (Table~\ref{tb:xmm}).
While the (unweighted) mean of the power law indices
derived from modelling the \xmm\ spectra is $\Gamma$$\sim$1.72, the mean index
at energies $\ga$2 keV is $\Gamma$$\sim$1.46 (Table~\ref{tb:xmm}),
close to that of the integrated CXRB spectrum.
This difference between observed and intrinsic spectral forms in red
AGN, which is similar to that seen in local, low luminosity AGN,
removes the discrepancy between the spectrum
of the CXRB and a subset of the AGN that may be 
responsible for it.

Thus in addition to Type 2 QSOs, there are two, potentially large, subsets
of the Type 1 population which
may contribute to the CXRB among AGN already known from their
optical/IR properties but not identified as X-ray hard in low S/N data:
those with significant X-ray
absorption, and those with undetected complexity.
A more detailed estimation of the possible CXRB contribution of red AGN
will be reported in a separate paper (Wilkes \etal\ 2004).

\subsection{Implications of the \xmm\ results for the nuclear regions of AGN.}

\xmm\ detects a soft X-ray emission component in all three Type 1 QSOs
in this sample.
Its form and relative luminosity, in all three cases, are similar to those
found in variability studies of the bright Seyfert 1 galaxies NGC4051 (Pounds
\et\ 2004a) and 1H0419-577 (Pounds \et\ 2004b), suggesting a common origin
in an extended region of moderately ionized gas. 
Its' detection in these sources with flatter
high-energy slopes, where a relatively weaker
power law would render a soft emission
component more visible, suggests it may be ubiquitous.

In Type 1 AGN, 2M2344, strong X-ray absorption is detected by \xmm\
while the broad optical lines remain visible at the usual strength.
This requires either that the (cold) absorber lies inside the BELR region
or is viewed at an intermediate line-of-sight so that the BELR remains
visible. A third alternative of purely scattered broad lines is unlikely
given the normal broad emission line strengths.
If interior to the BELR, such absorbing material must be
of high density to survive the intense ionising continuum.
To explain the different behavior of all three Type 1 QSOs
in this sample with a single scenario, 
an intermediate viewing angle through absorbing material in a
wind above/below the accretion disk/torus in current unification models
(Konigl \& Kartje 1994, Murray \& Chiang 1995,
Elvis 2000) can be combined with increased hot
dust emission and absorption.  The two Type 2 QSOs are then viewed at
a more edge-on orientation, so that the BELR is also obscured.
The presence of a large column of cold gas, in one/more components,
close to the Super-massive Black Hole (SMBH)
adds further complexity to the ``Unified
Scheme" which could help explain the classification-dependence of some AGN
on the waveband being used.

\section{Summary}

\xmm\ observations of five red AGN, selected as part of the 2MASS survey
by their red \jmk ($>2)$ color, show them to have a variety of
X-ray spectral forms. These results 
confirm and expand our earlier conclusions (Wilkes \etal\ 2002)
that these sources are X-ray hard. Their observed emission is hard at energies
above 2 keV even when it appears soft in the \chandra\ data.
This hardness is due either to absorption, as expected from
a simple obscuration scenario, or to a hard (intrinsic/reflected)
X-ray power law.
In several sources \xmm\ reveals complexity that was not apparent
in the \chandra\ data, emphasizing the limitations of low S/N data in
estimating X-ray absorption and/or spectral parameters.
Although the \chandra\ data can be misleading, the variety of behavior
results in no {\bf systemmatic}
effects on \chandra-deduced parameters which could be used to guide our
analysis of the larger sample.

The two Type 2 sources have the simplest spectra, showing absorption
$\sim 10^{22}$ cm$^{-2}$, at the low end of the distribution for
optical Type 2 AGN, and no further complexity. This is fully consistent with
the edge-on view expected for type 2 AGN in standard unification models.
All three Type 1 sources have complex spectra with a soft excess and some
combination of a normal power law, reflection and cold absorption.
The soft excess is best fit by a broad gaussian emission line which, in
this low S/N data, provides a reasonable approximation to the blend of
line emission expected from ionised gas.
It may be ubiquitous, though often hard to detect due to the presence of the
strong, unabsorbed, $\Gamma \sim 1.9$ intrinsic power law in most AGN.
It tends to be constant and it may originate in an extended
region of ionised gas (Pounds, Wilkes \& Page 2005).

The mean observed X-ray spectrum of these AGN at energies above 2 keV,
$\Gamma \sim 1.46$, matches well with that of the CXRB at these
energies. However, in all cases, the underlying power law has a normal,
$\Gamma = 1.7-1.9$ slope (Table~\ref{tb:xmm}). Thus within this red AGN
sample, there is no discrepancy between the observed
spectra of the AGN and the CXRB.
The combination of a hard observed X-ray spectrum above 2 keV and the
presence of broad emission lines in several cases suggests two types of
Type 1 QSOs which have been largely overlooked as contributors to the CXRB,
{\it i.e.} those with absorbed and/or complex spectra. Their
contributions may be significant. Red AGN are common at low-redshift, comprising
$\sim 20$\% of the population which is otherwise missed in current
optical surveys (Glikman \etal\ (2004).
Their numbers at higher redshift are less well-determined since the red
selection criterion is less effective at z$\ga 0.6$ (Cutri \etal\ 2002).
Improved information on their space density, in particular as a function
of redshift, will soon be provided by Spitzer-\chandra\ wide area surveys, 
allowing better estimates of both 
the importance of red AGN to the full AGN population 
and their contribution to the CXRB.

Significant X-ray absorption in Type 1 QSOs cannot be explained by simple
unification models. To cover the X-ray source but not the 
broad emission line region, the absorbing material must either be
small (nuclear) or patchy.
The intermediate level of the absorption, $10^{22}$ cm$^{-2}$,
is suggestive of the outflowing wind/atmosphere above a
disk/torus (Konigl \& Kartje 1994, Murray \& Chiang 1995)
in current unification models
and thus of an intermediate viewing angle for red AGN.
This scenario can also explain the observed red optical and near-IR colors.
However, it can only explain the properties of all three Type 1
if we also invoke unusually high levels of hot dust emission
and/or absorption.
Thus X-ray observations of these
red AGN will not only provide a check on the
presence of absorption, but also a unique probe of
the absorbing material. Although X-ray variability, which generally occurs
on a much faster timescale
than seen in the near-IR for radio-quiet AGN such as these
(Neugebauer \& Matthews 1999), 
may make it hard to find a consistent picture.

\section{Acknowledgments}

The results reported here are based on observations obtained with \xmm, an ESA science
mission with instruments and contributions directly funded by ESA Member States and the USA (NASA). The authors wish to
thank the SOC and SSC teams for organising the \xmm\ observations 
and initial data reduction. KAP is pleased to acknowledge a Leverhulme Trust Emeritus Fellowship.
BJW and GDS are grateful for the financial support of \xmm\ GO grant: NNG04GD27G.


\clearpage

\begin{table}
\begin{center}
\caption{XMM-Newton Observation details}
\label{tb:obs}
\begin{tabular}{llccclcc}
&\\

\tableline
 &  \multicolumn{3}{c}{pn camera} & & \multicolumn{3}{c}{MOS cameras} \\
 \cline{2-4} \cline{6-8}\\
2MASS J  & filter & exposure (s) & counts & &filter & exposure (s) & counts\\
\tableline

09184860+2117170 & thin & 5439 & 288$\pm$60 & &thin & 16098 & 241$\pm$32\\
10514425+3539306 & thin & 3044 & 1126$\pm$61 & &medium & 10645 & 1490$\pm$43\\
13000534+1632149 & medium & 1184 & 284$\pm$59 & &medium & 5355 & 396$\pm$32\\
14025120+2631175 & medium & 1699 & 1648$\pm$51 & &medium & 4937 & 1136$\pm$44\\
23444958+1221432 & thin & 4444 & 844$\pm$73 && thin & 13164 & 790$\pm$38 \\

\tableline
\end{tabular}

\end{center}
\end{table}

\begin{table}[h]
\begin{center}
\caption{\chandra\ X-ray and Optical Parameters }
\label{tb:Xopt}
\begin{tabular}{lllcclccr}
&\\
\tableline
2MASS J  & Redshift & $\Gamma$$^(1)$ & \nh$_{int}$ & HR & \jmk\ & B$-$R &Class$^{2}$ & \%P$^{3}$ \\
&&& /$10^{21}$cm$^{-2}$ &  & & (DSS) &&\\\tableline
0918 & 0.149 & 1.9$\pm$0.5 & 2$\pm2$ & $-0.6 \pm 0.1$ & 2.23$\pm$.04 & 2.1 & 1.5 & 10.01$\pm$0.07\\
1051 & 0.158 & 1.6$\pm$0.2 & 5.6$\pm1.4$ & $-0.2\pm 0.1$ & 2.08$\pm$.07
& 2.6 & 1.9 & 0.54$\pm$0.29 \\
1300 & 0.080 & 1.5$\pm$0.8 & 18$\pm$10 & $+0.2\pm 0.1$  & 2.17$\pm$.05  &
3.1 & 2.0 & 9.50$\pm$0.07 \\
1402 & 0.187 & 2.0$\pm$0.4 & $0.1\pm1.3$ & $-0.6\pm 0.1$ & 2.08$\pm$.05 &
0.6 & 1.0 & 0.21$\pm$0.21\\
2344 & 0.199 & 2.0$\pm$0.4 & 4$\pm$2 & $-0.4\pm 0.1$ & 2.00$\pm$.06 &
1.3 & 1.0 & 1.01$\pm$0.24 \\
\tableline
\end{tabular}
\end{center}
\begin{minipage}{6.5in}
1: Power law plus absorption fit over 0.5--8 keV band \\
2: Optical class: 1.0$-$2.0, based on the ratio of narrow to broad
emission lines \\
3: Optical \% polarisation, Smith \etal 2002, 2003 \\
\end{minipage}
\end{table}

\clearpage
\begin{table}
\begin{center}
\caption{\xmm\ X-ray Spectral Parameters (pn camera)}
\label{tb:xmm}
\begin{tabular}{llccccc}
&\\
\tableline
2M & $\Gamma$(2-10) & $\Gamma$(0.3$-$10, pn) & \nh &\multicolumn{2}{c}{Line} & $\chi^2$/dof   \\
&&&/$10^{21}$cm$^{-2}$ & energy(keV) & flux$^{1}$  \\\tableline
0918 &1.26$\pm$0.39 &1.65$\pm$0.33$^2$ & $< 1.1^3$ & 0.44$\pm$0.13 & $4\times10^{-5}$ & 13/20
 \\
1051 & 1.67$\pm$0.11&1.79$\pm$0.08 & 7.2$\pm$0.7 & - & - &103/115 \\
1300 &1.16$\pm$0.16 &1.74$\pm$0.26 & 29$\pm$5 & 6.2 &$1.3\times10^{-5}$ & 11/24
\\
1402 & 1.53$\pm$0.16&1.54$\pm$0.07 & $< 0.4^3$ & 0.44$\pm$0.10 & $4\times10^{-4}$ & 110/102
\\
2344 & 1.70$\pm$0.16&1.86$\pm$0.12 & 6.5$\pm$2.6 & 0.64$\pm$0.05 & $1\times10^{-4}$ & 65/69
 \\

\tableline
\end{tabular}
\end{center}
\begin{minipage}{6.5in}
1: photon cm$^{-2}$ s$^{-1}$ \\
2: The best fit model includes a cold reflection component: R$\sim 5$
and EW (FeK$\alpha$)$\sim 0.5$ keV (at 6.4 keV) \\
3: 90\% upper limit. When absorption at this level is included, the best
fit power law becomes: $\Gamma =1.8$ (2M0918), $\Gamma =1.64$ (2M1402).
\end{minipage}
\end{table}

\clearpage
\begin{table}
\begin{center}
\caption{\xmm\ X-ray Fluxes and Luminosities}
\label{tb:flux}
\begin{tabular}{lccccc}
&\\
\tableline
2M & Flux(0.3-1 keV)$^1$ & Flux(1-2 keV)$^1$ & Flux(2-10 keV)$^1$ & L(0.3-10 keV)$^2$  & L$_{SX}^3$
 \\
& /$10^{-13}$ & /$10^{-13}$ & /$10^{-12}$ & /$10^{44}$ & /$10^{42}$  \\\tableline
0918 & 0.41 & 0.32 & 0.19 & 0.14 & 0.6   \\
1051 & 0.5 & 4 & 2.0 & 2.3 & - \\
1300 & 0.14 & 1.45 & 2.35 & 0.68 & - \\
1402 & 7.5 & 4.7 & 1.9 & 2.6 & 16 \\
2344 & 0.9 & 1.4 & 0.51 & 0.79 & 2.9 \\

\tableline
\end{tabular}
\end{center}
\begin{minipage}{6.5in}
1: erg cm$^{-2}$ s$^{-1}$ \\
2: Absorption-corrected luminosity in units of erg s$^{-1}$  \\
3: Absorption-corrected luminosity of the soft excess component in erg s$^{-1}$ \\
\end{minipage}
\end{table}

\clearpage
\begin{figure}
\centering
\rotatebox{-90}{
\epsscale{1.2}
\plottwo{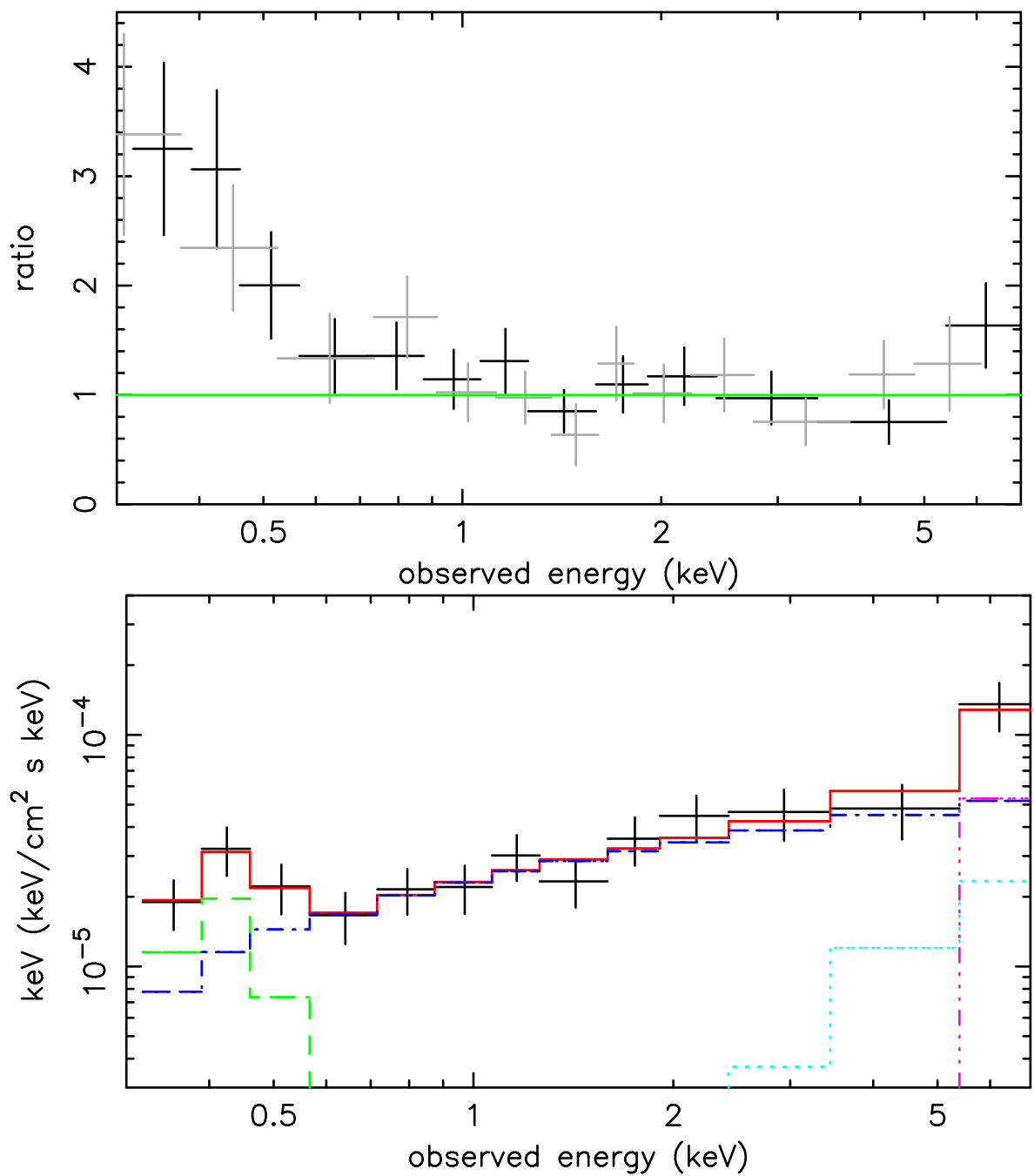}{f2.eps}}
\caption{2MASS J091848+2117 a(upper): Ratio of pn (black) and MOS (grey) EPIC spectral data
to a hard power law fit above 2 keV revealing a soft excess
and evidence for reflection in the highest energy channels.
b(lower): Unfolded spectrum (black) illustrating the best-fit
(red), with components: power law (dark blue) plus reflection (light
blue), Fe K$\alpha$ line (magenta) and soft emission component (green).
Only pn camera data are shown for clarity.
\label{fig1}}
\end{figure}


\clearpage

\begin{figure}
\rotatebox{-90}{
\epsscale{1.2}
\plottwo{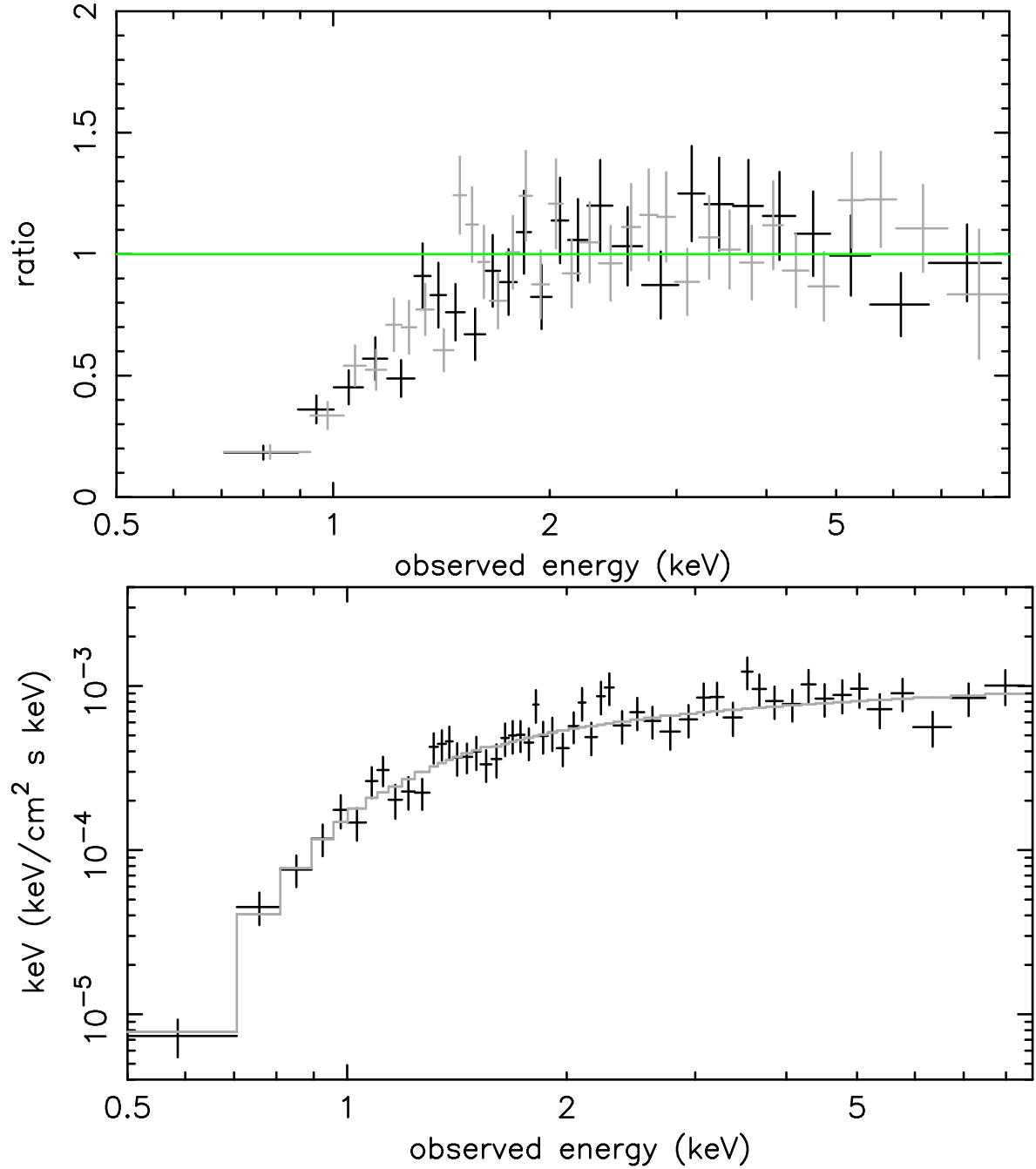}{f4.eps}}
\caption{2MASS J105144+3539 a(upper): Ratio of pn (black) and MOS (grey) EPIC
spectral data to a simple power 
law fit above 2 keV, showing a strong low energy cut-off. 
b(lower): Unfolded spectrum (black) illustrating the best-fit attenuated power law model 
(grey). Only pn camera data are shown for clarity.
\label{fig2}}
\end{figure}


\clearpage

\begin{figure}
\rotatebox{-90}{
\epsscale{1.2}
\plottwo{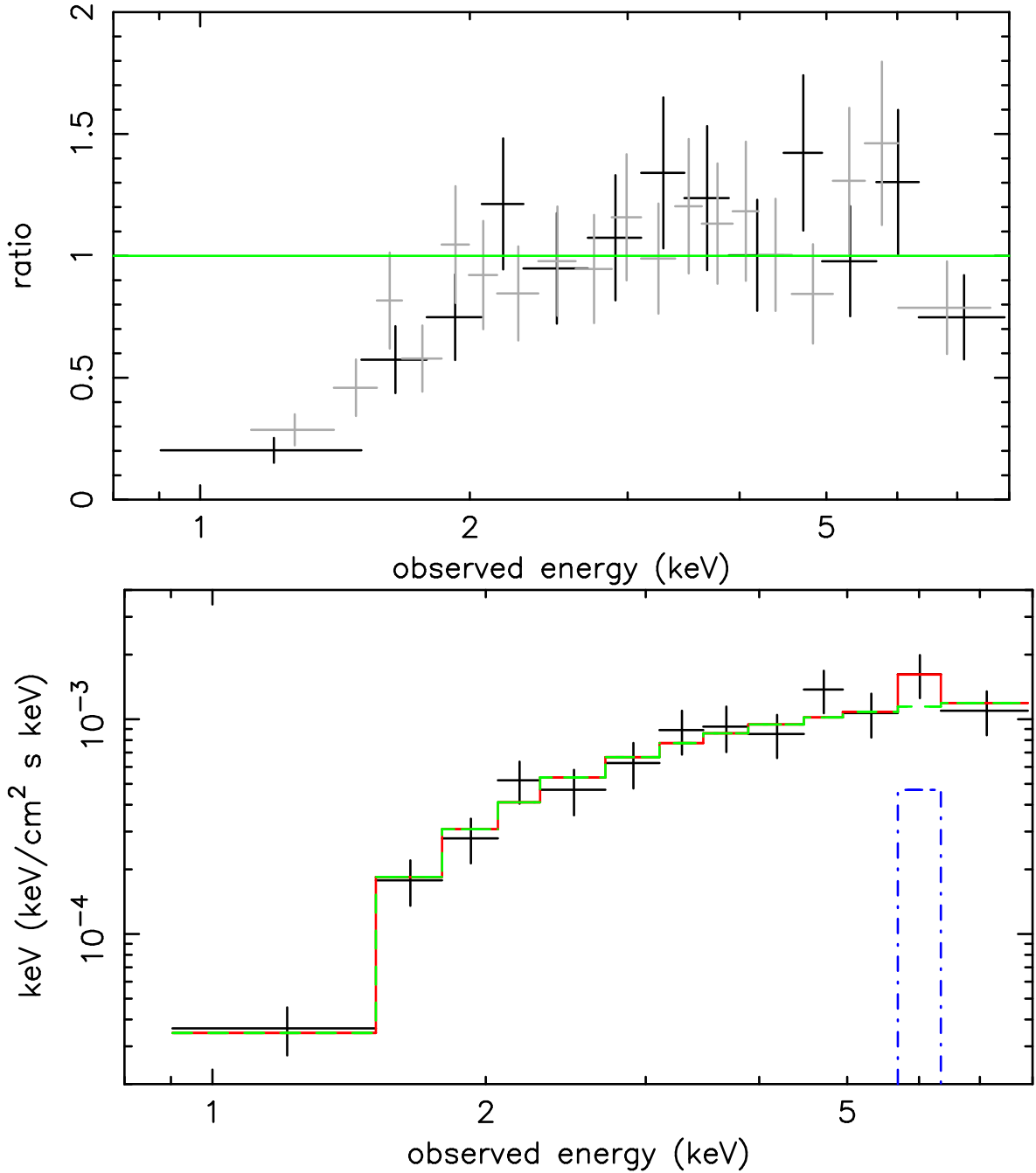}{f6.eps}}
\caption{2MASS J130005+1632 a(upper): Ratio of pn (black) and MOS (grey) EPIC spectral data
to a simple power law fit above 2 keV, showing a strong low energy cut-off. 
b(lower): Unfolded spectrum (black) illustrating the best-fit (red)
attenuated power law model (green). A weak Fe K emission line (blue) is seen near 6 keV.
Only pn camera data are shown for clarity.
\label{fig3}}
\end{figure}


\clearpage

\begin{figure}
\rotatebox{-90}{
\epsscale{1.2}
\plottwo{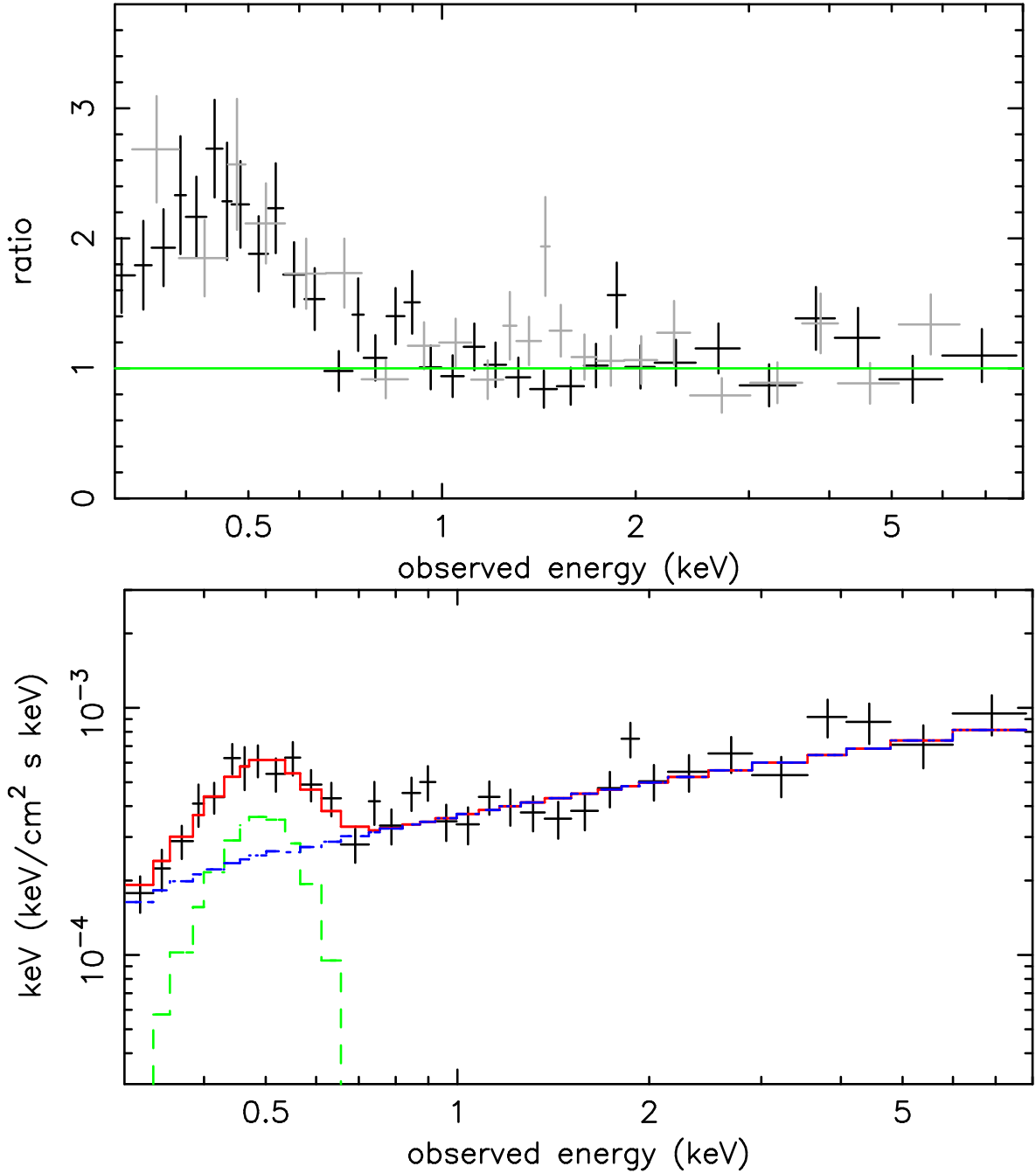}{f8.eps}}
\caption{2MASS J140251+2631 a(upper): Ratio of pn (black) and MOS (grey) EPIC spectral data
to a simple power law fit above 2 keV, showing a soft excess when extrapolated to
lower energies. 
b(lower): Unfolded spectrum (black) illustrating the best-fit model
(red), with components: power law (blue) plus gaussian soft emission component
(green). Only pn camera data are shown for clarity.
\label{fig4}}
\end{figure}


\clearpage

\begin{figure}
\rotatebox{-90}{
\epsscale{1.2}
\plottwo{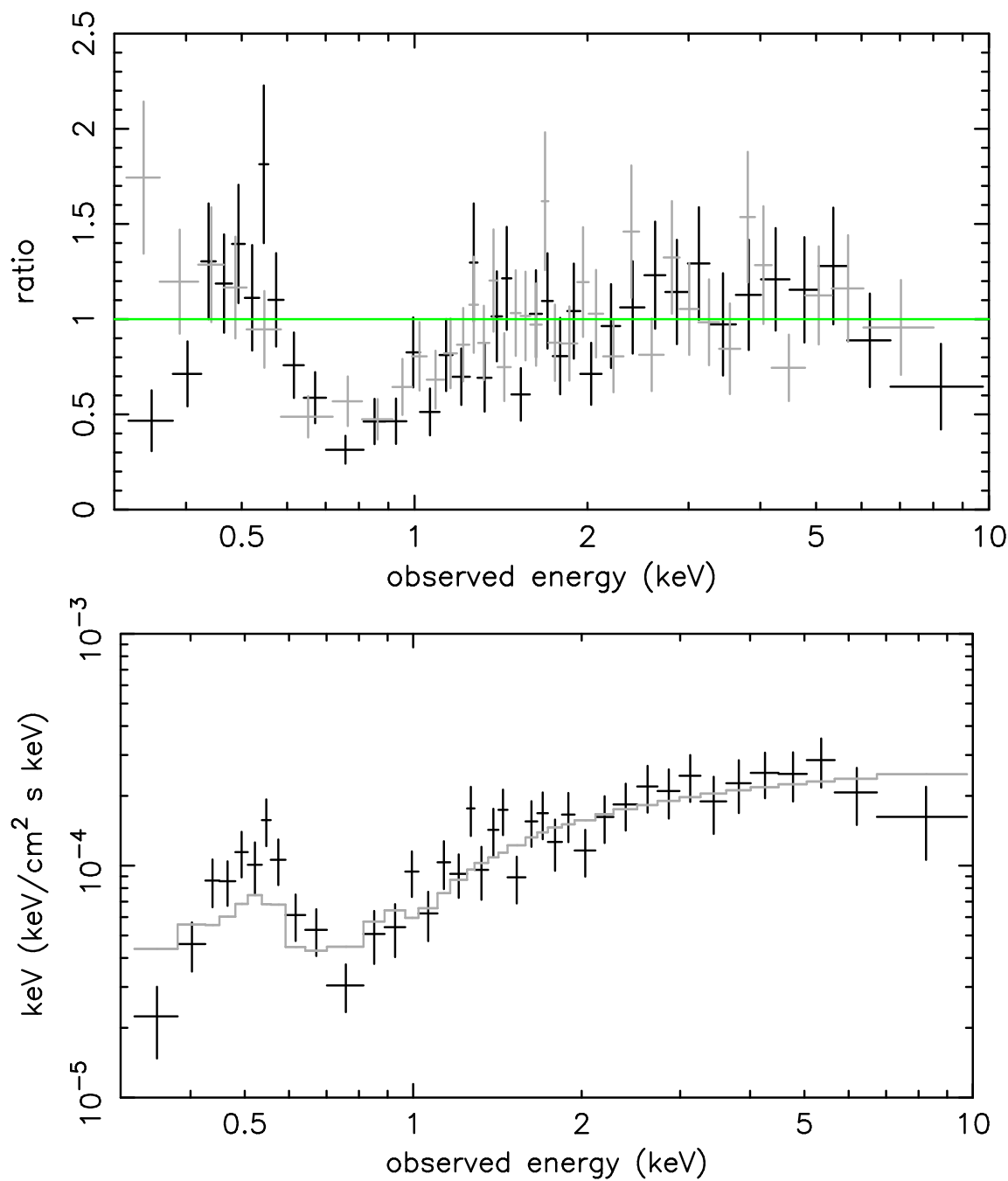}{f10.eps}}
\caption{2MASS J234449+1221 a(upper): Ratio of pn (black) and MOS (grey) EPIC spectral data
to a simple power law fit above 2 keV, showing evidence for low energy absorption and a
possible soft excess. b(lower): Unfolded spectrum (black) illustrating the best-fit power
law plus photoionised absorber model (grey). Only pn camera data are shown for clarity.
\label{fig5}}
\end{figure}



\begin{figure}
\rotatebox{-90}{
\epsscale{0.6}
\plotone{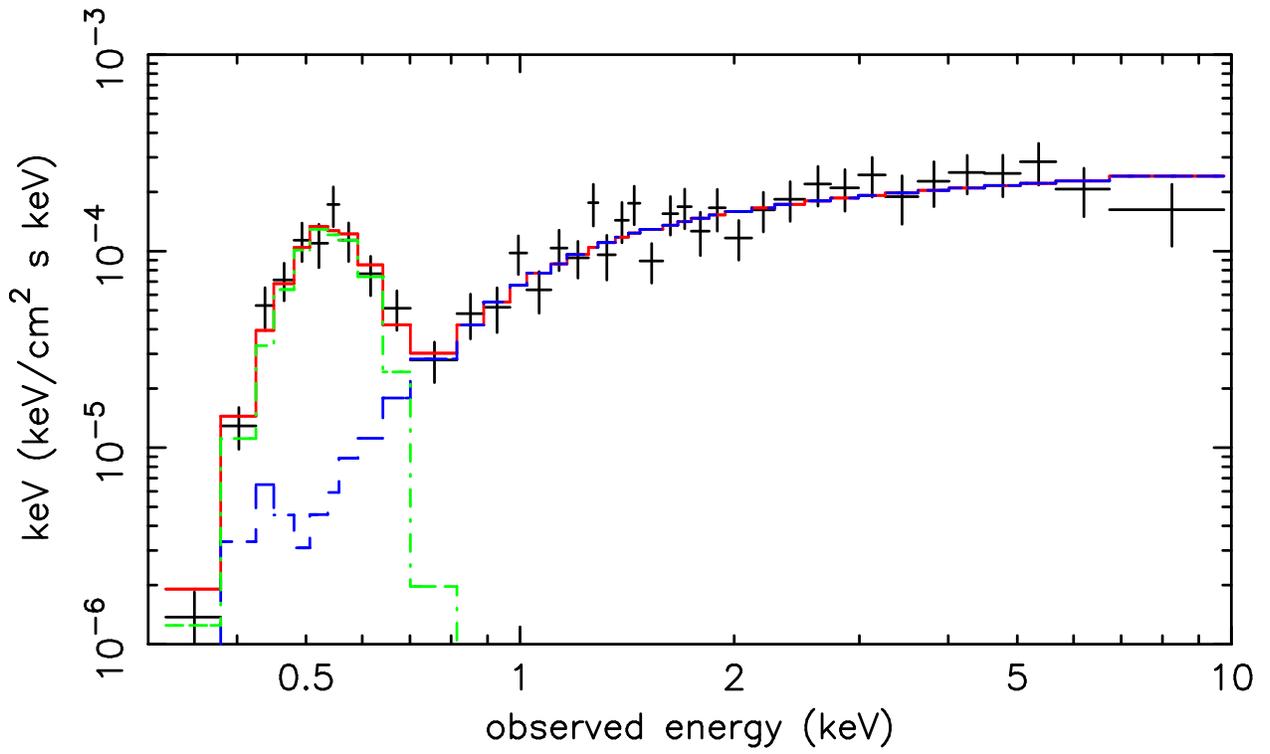}}
\caption{2MASS J234449+1221: Unfolded spectrum (black) illustrating the best-fit
(red) absorbed power law (blue) plus gaussian emission component
(green) model.
Only pn camera data are shown for clarity.
\label{fig6}}
\end{figure}

\end{document}